\documentclass[prd, twocolumn, nofootinbib, floatfix]{revtex4-1}

\usepackage{amsmath}
\usepackage{graphicx}
\usepackage{dcolumn}
\usepackage{bm}
\usepackage{epsfig}
\usepackage{amssymb,latexsym,mathrsfs}
\usepackage{graphicx}
\usepackage{color}
\usepackage{hyperref}

\hypersetup{
    colorlinks=true,
    linkcolor=red,
    citecolor=blue,
} 

\newcommand{\be}{\begin{equation}}
\newcommand{\ee}{\end{equation}}
\newcommand{\beq}{\begin{equation}}
\newcommand{\eeq}{\end{equation}}

\newcommand{\bs}{\begin{split}} 
\newcommand{\bea}{\begin{eqnarray}}
\newcommand{\eea}{\end{eqnarray}}
\newcommand{\beqa}{\begin{eqnarray}}
\newcommand{\eeqa}{\end{eqnarray}}

\newcommand{\geff}{G_{\rm eff}} 
\newcommand{\geffds}{G_{\rm eff,dS}} 
\newcommand{\geffh}{G_{\rm eff}^{(\phi)}} 
\newcommand{\geffs}{G_{\rm eff}^{(\psi)}} 
\newcommand{\mpl}{M_{\rm Pl}}

\begin{document}

\title{Are Scalar and Tensor Deviations Related in Modified Gravity?} 
\author{Eric V.\ Linder} 
\affiliation{Berkeley Center for Cosmological Physics \& Berkeley Lab, 
University of California, Berkeley, CA 94720, USA} 

\begin{abstract}
Modified gravity theories on cosmic scales have three key deviations 
from general relativity. They can cause cosmic acceleration without a 
physical, highly negative pressure fluid, can cause a gravitational slip 
between the two metric potentials, and can cause gravitational waves to 
propagate differently, e.g.\ with a speed different from the speed of light. 
We examine whether the deviations in the metric potentials as observable 
through modified Poisson equations for scalar density perturbations are 
related to or independent from deviations in the tensor gravitational waves. 
We show analytically they are independent instantaneously 
in covariant Galileon gravity -- e.g.\ at some time 
one of them can have the general relativity value while the other deviates -- 
though related globally -- if one deviates over a finite period, the 
other at some point shows a deviation.  We present expressions for the 
early time and late time de Sitter limits, and numerically illustrate 
their full evolution. 
This in(ter)dependence of the scalar and tensor deviations highlights 
complementarity between cosmic structure surveys and future gravitational 
wave measurements. 
\end{abstract}

\date{\today} 

\maketitle

\section{Introduction} 

Extensions to general relativity have become a subject of intense interest 
in the last decade, due to both observational evidence for acceleration of 
the cosmic expansion and intriguing new theoretical work. For general 
relativity to explain cosmic acceleration it requires a physical 
energy-momentum component that violates the strong energy condition, for 
example a cosmological constant or scalar field with highly negative 
pressure. These explanations have difficulties with fine tuning and 
naturalness, so an attractive alternative has been to consider altering 
the structure of the gravitational action itself, e.g.\ through 
scalar-tensor theories, higher dimensional gravity, or massive gravitons. 

Inventing a sound, consistent gravity theory is no easy task, and furthermore 
the theory must satisfy observational cosmology constraints such as an early 
universe behavior similar to general relativity, with radiation and matter 
domination, a deviation near the present to explain cosmic acceleration, 
and growth of large scale massive structures not too dissimilar from in 
general relativity. Indeed the specific expansion and growth histories, 
and their comparison, provide one of the key signatures of modified gravity. 
The evolution of linear density perturbations, and their gravitational 
lensing of light, can be phrased in terms of, respectively, nonrelativistic 
and relativistic modified Poisson equations where Newton's constant becomes 
two distinct functions of space and time. These effective gravitational 
strengths can not only differ from Newton's constant, but from each other; 
this divergence is called the gravitational slip. 

Observationally there are no constraints yet on cosmological gravitational 
waves, but many modified theories predict that their propagation will differ 
from general relativity; 
in particular, as we focus on here, the sound speed of these tensor 
perturbations may deviate from the speed of light. 
Recently, 
\cite{14067139} illustrated a relation between the gravitational slip 
and the gravitational wave propagation in several modified gravity theories: 
a deviation in one led to a deviation in the other. 

Here we consider the generality of such a relation. If it arose from the 
intrinsic structure of the gravity theory, it would be a powerful tool for 
detecting some modifications and predicting others. We work within 
Horndeski gravity, the most general single scalar field gravity that obeys 
second order field equations, and in particular Galileon gravity plus an 
extension with disformal couplings. 

In Sec.~\ref{sec:gali} we review the relevant equations of motion that lead 
to the modified Poisson equations defining the gravitational strength 
functions and their slip, as well as the sound speed of gravitational waves. 
Section~\ref{sec:desitter} examines these expression analytically, in the 
early time limit and for the asymptotic de Sitter late time attractor of 
cosmic acceleration. We display the full evolution for various cases in 
Sec.~\ref{sec:num} and discuss the results in Sec.~\ref{sec:concl}.

\section{Modifications beyond General Relativity} \label{sec:gali} 

The metric for a spatially flat Friedmann-Robertson-Walker cosmology 
linearly perturbed by scalar density modes is 
\beq 
ds^{2} = -(1+2\psi) dt^{2} + a^{2} (1-2\phi) \delta_{ij}dx^{i}dx^{j} \,, 
\eeq 
in Newtonian gauge, where $a$ is the cosmic expansion factor, and 
$\psi$ and $\phi$ are the metric potentials. 
We will be particularly interested in whether they are equal, as in general 
relativity (we assume no fluid anisotropic stress, which can generate a 
difference between them). The gravitational slip is defined as 
\beq 
\eta=\frac{\phi}{\psi} \,, 
\eeq 
and we will explore the conditions under which it deviates from unity 
in a modified gravity theory. 

The metric potentials can be probed observationally by the growth of 
linear density perturbations and the gravitationally lensing they induce. 
For subhorizon scales we write the modified Poisson equations in the 
standard quasistatic approximation (see Appendix~\ref{sec:apxalf} 
and \cite{11084242} for 
discussion of the quasistatic approximation in unusual cases) as 
\beqa 
& & \nabla^{2} \phi = 
4\pi a^{2} G^{(\phi)}_{\rm eff}\rho_m \,\delta_m  \\  
& &   \nabla^{2} \psi = 4\pi a^{2} G^{(\psi)}_{\rm eff} \rho_m\,\delta_ m \\  
& &   \nabla^{2} (\psi+\phi) = 8\pi a^{2} G^{(\psi+\phi)}_{\rm eff} 
\rho_m\,\delta_m \,. 
\eeqa 
Here the $\geff$ are the modified Newton's constants giving the gravitational 
strength, $\rho_m$ is the matter density, and $\delta_m=\delta\rho_m/\rho_m$ 
is the density perturbation. 
The second equation is central to the motion of nonrelativistic 
matter and hence the growth of massive structures, while the third is 
central to the motion of relativistic particles, and hence the propagation 
of light \cite{bert11}. Note that $\geff^{(\psi+\phi)}=\geffh+\geffs$. 
The gravitational slip is then 
\beq 
\eta=\frac{\geffh}{\geffs} \ . 
\eeq 

We now specialize to covariant Galileon gravity \cite{gal1,gal2,gal3}, a 
subset 
of the general Horndeski theory \cite{horndeski,deffayet,kobayashi,11084242}, 
except that we will 
also allow a derivative coupling term related to 
the increasingly investigated disformal field theories 
\cite{amen93,disformal,disformal2,disformal3}. Such a theory has a number of 
useful 
properties, such as involving only second order equations of motion and 
having a shift symmetry (as well as softly broken Galilean symmetry). 

The action following the notation of \cite{appgal} is 
\beqa 
S =&& \int d^4 x\,\sqrt{-g} \left[ \frac{M_{\rm Pl}^2}{2} R - \frac{c_{2}}{2} \right.
(\partial \pi)^{2} - \frac{c_{3}}{M^{3}}(\partial \pi)^{2} \Box \pi 
\notag\\ 
& & -\frac{c_{4}}{2}{\cal L}_{4} - \frac{c_{5}}{2}{\cal L}_{5} 
- \frac{M_{\rm Pl}}{M^{3}} c_{\rm G} G^{\mu\nu} 
\left. \partial_{\mu}\pi\partial_{\nu}\pi  - {\cal L}_{m} \right] \,, 
\eeqa 
where $\pi$ is the Galileon scalar field, the $c_i$ are constant 
coefficients, ${\cal L}_m$ is the fluid Lagrangian, and 
\beqa 
{\cal L}_{4} = & &\ (\nabla_{\mu}\pi)(\nabla^{\mu}\pi) 
\left[ 2(\Box\pi)^{2} - 2 \pi_{;\mu\nu}\pi^{;\mu\nu} \right.\notag\\ 
& & \left. - R (\nabla_{\mu}\pi)(\nabla^{\mu}\pi)/2 \right]/M^{6}\\ 
{\cal L}_{5} = & &\ (\nabla_{\mu}\pi)(\nabla^{\mu}\pi)
\left[ (\Box\pi)^{3} - 3(\Box \pi)\pi_{;\mu\nu}\pi^{;\mu\nu} \right.\notag\\ 
& & \left. + 2 \pi_{;\mu}{}^{;\nu}\pi_{;\nu}{}^{;\rho}\pi_{;\rho}{}^{;\mu} - 6\pi_{;\mu}\pi^{;\mu\nu}\pi^{;\rho}G_{\nu\rho} \right] /M^{9} \,. 
\eeqa 

The equations of motion can be written as coupled first order differential 
equations for the Hubble expansion $H=d\ln a/d(H_0t)$ and field evolution 
$x=d(\pi/M_{\rm Pl})/d\ln a$, as in \cite{appgal}. The gravitational slip 
becomes 
\beq 
\eta = \frac{\kappa_4 \kappa_6 -\kappa_5 \kappa_1}{2(\kappa_3 \kappa_6 - \kappa_1^2)} \,, \label{eq:eta} 
\eeq 
where the $\kappa_i$ involve sums of terms depending on $c_i$, $H$, and $x$ 
(see Appendix~\ref{sec:apxkap} for the explicit expressions). When all $c_i=0$, i.e.\ 
general relativity, then all $\kappa_i$ vanish except $\kappa_3=-1$ and 
$\kappa_4=-2$, so in this case indeed $\eta=1$. 

The expression for the gravitational wave speed $c_T$ for Horndeski gravity 
appears in Eq.~(28), with Eqs.~(17) and (20), of \cite{1110.3878}. 
Note that for the Galileon gravity we consider, the relevant 
Horndeski functions 
\beqa 
& & G_4=\frac{1}{2}+\frac{c_4}{4}H^4 x^4\\ 
& & G_5=\frac{-3c_5}{4}H^4 x^4 +c_G\frac{\pi}{\mpl} \,. 
\eeqa 
The $c_G$ term comes from the derivative coupling to the Einstein tensor, 
$c_G G^{\mu\nu}\nabla_\mu\pi \nabla_\nu\pi$, we have included in the 
Galileon action. This disformal coupling also allows us to study other 
gravity theories of interest such as purely kinetic coupled gravity 
\cite{gubitosi} and some aspects of Fab 4 \cite{fab4} and Fab 5 gravity 
\cite{fab5}. 

The tensor sound speed is then given by 
\beqa 
c_T^2=& &\ \frac{2\kappa_3}{\kappa_4} \label{eq:ct}\\ 
=& &\ \frac{1+\frac{c_4}{2}H^4 x^4+3c_5 H^6 x^5\left(\frac{H'}{H}+\frac{x'}{x}\right) -c_G H^2 x^2}{1-\frac{3c_4}{2}H^4 x^4+3c_5 H^6 x^5+c_G H^2 x^2} \ ,\notag 
\eeqa 
where prime denotes $d/d\ln a$. 
Note that $c_2$ and $c_3$ do not explicitly enter; this is an important point. 

If only $c_2$ and $c_3$ are nonzero, then $\kappa_1$ vanishes, 
$\kappa_4=2\kappa_3$, and then from Eq.~(\ref{eq:eta}) we have $\eta=1$. 
This is also apparent from the full equation for $\phi-\psi$ in 
Eq.~(C1) of \cite{appgal}. This raises the conjecture that there are some 
conditions under which the gravitational wave speed and gravitational 
slip are closely connected, as investigated by \cite{14067139}. 

However, 
we see that $c_4$, $c_5$, and $c_G$ can all contribute to both $c_T$ and 
$\eta$; if any of them are nonzero then there can be deviations from 
general relativity in these observables. It is not obvious that these 
terms enter these two quantities in the same way, though, and so the potential 
exists for a deviation signature to be evident in only one of them. That is, 
we want to see if we can have $\eta=1$ but $c_T\ne1$, or $c_T=1$ but 
$\eta\ne1$, or if indeed a deviation in one forces a deviation in the other.

\section{Early and Late Time Limits} \label{sec:desitter} 

In the early radiation and matter dominated eras we generally want 
modified gravity deviations from general relativity to be small, to 
preserve excellent agreement with primordial nucleosynthesis and cosmic 
microwave background observations. For the Galileon case, \cite{appgal} 
calculated that at early times 
$\geffh=1+{\cal O}(\Omega_\pi)$, where $\Omega_\pi$ is 
the fraction of critical density contributed by the modified gravity 
scalar field. We can carry out a similar computation for $\geffs$ and 
find that also $\geffs=1+{\cal O}(\Omega_\pi)$; for example if the $c_5$ 
term dominates the Galileon Lagrangian as expected at early times then 
$\geffs=1+(759/224)\Omega_\pi$ during matter domination. 

Thus $\eta_{\rm early}=1+{\cal O}(\Omega_\pi)$, and we can similarly 
calculate that $c^2_{T,{\rm early}}=1+{\cal O}(\Omega_\pi)$. 
For example, in matter domination with the leading $c_5$ term we have 
\beqa 
& & \eta_{\rm early}=1+\frac{111}{32}\Omega_\pi \\ 
& & c^2_{T,{\rm early}}=1+\frac{15}{56}\Omega_\pi \ . 
\eeqa 
Thus at early times 
a deviation signature in one, while small, does imply a deviation in the 
other, since they both arise from the effective dark energy density (since 
only one term in the Lagrangian dominates). 

At later times, however, multiple Lagrangian terms can be comparable and 
this connection can be broken. We show the evolution of $\eta$ and $c_T$ 
in the next section, but first we demonstrate analytically the breakdown 
of the connection between the scalar and tensor deviations for late time 
cosmology, i.e.\ when the effective dark energy is 
non-negligible, and in particular when it completely dominates in the 
de Sitter limit. 

Let us define 
\beq 
e=c_4 H^4 x^4 \,, \ \ f=c_5 H^6 x^5 
\,, \ \ a=c_G H^2 x^2 \,. \label{eq:efa} 
\eeq 
We can write 
\beqa
c_T^2-1=& &\ \frac{2e-2a-3f\left(1-\frac{H'}{H}-\frac{x'}{x}\right)}{1-\frac{3}{2}e+a+3f} \label{eq:ct21}\\ 
\kappa_1\,x=& &\ -(c_T^2-1)\left(1-\frac{3}{2}e+a+3f\right)-6e\left(\frac{H'}{H}+\frac{x'}{x}\right)\notag\\ 
& &\ +2a\left(\frac{H'}{H}+\frac{x'}{x}\right) 
+3f\left(\frac{4H'}{H}+\frac{3x'}{x}\right)\\ 
\kappa_4=& &\ 2\kappa_3+2(c_T^2-1)\left(1-\frac{3}{2}e+a+3f\right) \ . 
\eeqa 
From Eq.~(\ref{eq:eta}) we see that 
\beq 
\eta-1 \propto (\kappa_4-2\kappa_3)\kappa_6+\kappa_1(2\kappa_1-\kappa_5) \ , 
\label{eq:eta1} 
\eeq 
so to achieve vanishing slip at some moment the vanishing deviation $c_T^2-1$ 
is insufficient -- one must also have 
the conditions $H'=0=x'$ (or very small $\kappa_1$ as in the early 
universe when it is proportional to $\Omega_\pi$, or an instant of 
cancellation in the evolution of the various terms from the Lagrangian). 

The conditions $H'=0=x'$ are the fixed points for the equations of motion, 
corresponding to de Sitter cosmology. In fact, in the de Sitter limit 
$\eta_{\rm dS}=1$ regardless of the value of $c_T^2$, as noted in 
\cite{appgal}. (We can see this here by substituting the algebraic 
constraints Eqs.~(67)--(68) from \cite{appgal} into Eq.~(\ref{eq:eta1}), 
resulting in its right hand side vanishing.) But in general, except for 
these two exceptions -- de Sitter late time limit and negligible effective 
dark energy early time limit -- $c_T^2=1$ at some moment does not imply 
$\eta=1$ then. Nor is the converse true: $\eta=1$ at some moment does not 
imply $c_T^2=1$ then, except at early times (not even in the de Sitter limit, 
as we consider below). Basically, the two equations of motion give two 
constraints, and fixing the present dark energy density would give a third, 
but this is only three equations for 
five quantities ($c_{2\dots 5}$, $c_G$) and so $\eta$ is not completely 
determined without further assumptions about the values of the $c_i$; 
in particular nothing forces $\eta=1$. So $c_T^2=1$ does not guarantee 
$\eta=1$. 

(One might note that tensor deviations can occur in other ways 
than through changing the gravitational wave speed, i.e.\ through a 
graviton mass term, a transverse traceless source term, or a running of 
the Planck mass \cite{14067139}. 
These potentially add more parameters, which further underdetermines the 
system. However, in theories where these extra effects are independent 
from the tensor sound speed \cite{14043713} and if they are fixed to zero 
deviation from general relativity, these extra conditions might be sufficient 
to close the system and guarantee $\eta=1$. This is what occurs in 
the Horndeski case considered by \cite{14067139}; we discuss it further in 
Appendix~\ref{sec:apxalf}. 
However, this does not help much since by Eq.~(\ref{eq:eta1}) this forces a 
constraint on the expansion history $H$, and does not solve the following 
counterexample.) 

Now consider the converse: does $\eta=1$ imply $c_T^2=1$? This obviates 
the issue of tensor mode propagation depending on more than the gravitational 
wave sound speed. If lack of deviations from general relativity in the 
scalars must imply lack of deviations in the tensor modes, then $c_T^2=1$ 
is a necessary condition. 

One cannot obtain a de Sitter state with only one of the $c_i$'s nonzero. 
Since we require nonzero $c_4$, $c_5$, or $c_G$ to have $c_T^2\ne1$, for simplicity 
we could look at pairs involving at least one of them. Indeed we then find 
that these cases violate $c_T^2=1$. However, although this shows that 
nothing in the structure of the theory requires scalar mode deviations 
(or lack thereof) to guarantee tensor mode deviations (or lack thereof), 
these ``pair'' cases do generally  have an instability in the scalar sound 
speed, $c_s^2<0$, or sometimes a ghost (see Fig.~5 in \cite{appgal}) and so 
one might prefer a purely 
healthy theory. Therefore we take the simple illustrations of some triplet 
cases. 

Recall that $\eta_{\rm dS}=1$. It is convenient to define $E$, $F$, $A$ as the 
values that $e$, $f$, $a$ from Eqs.~(\ref{eq:efa}) 
take at the de Sitter fixed point. We then find 
\beqa 
& & c_{T,{\rm dS}}^2=\frac{1+\frac{1}{2}E}{1-\frac{3}{2}E} \quad{\rm for\ }\{c_2,c_3,c_4\} \\ 
& & c_{T,{\rm dS}}^2=\frac{1}{1+3F} \quad{\rm for\ } \{c_2,c_3,c_5\} \\ 
& & c_{T,{\rm dS}}^2=\frac{1-A}{1+A} \quad\ {\rm for\ }\{c_2,c_3,c_G\} \ . \label{eq:ctcg} 
\eeqa 
This shows that any of $c_4$, $c_5$, $c_G$ can give deviations in the tensor 
sector while keeping the scalar slip as general relativity. 
The last case is particularly interesting since the derivative coupling 
shows up in purely kinetic gravity \cite{gubitosi}, Fab 4 gravity \cite{fab4}, 
and Fab 5 gravity \cite{fab5}. Indeed Eq.~(\ref{eq:ctcg}) appears in 
Eq.~(B23) of \cite{fab5}, using the opposite sign convention for $c_G$. 
(Recall that $c_2$ and $c_3$ do not contribute 
explicitly to $c_T^2$ and so are somewhat irrelevant.) 

In the next section we exhibit numerical solutions where all terms are 
present, at arbitrary times. Finally we consider the global situation 
where one sector looks like general relativity at all times.

\section{Evolution of Deviations} \label{sec:num} 

\begin{figure}[htbp!]
\includegraphics[width=\columnwidth]{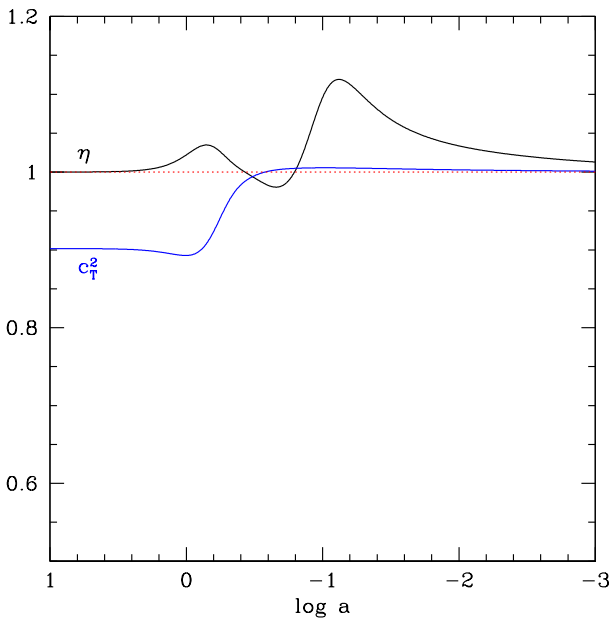} 
\caption{The evolution of the gravitational slip $\eta$ and gravitational 
wave sound speed squared $c_T^2$ is plotted vs the log of the expansion 
factor for an uncoupled Galileon gravity model. At late times the slip 
goes to unity, the general relativity value, but $c_T^2$ does not. 
} 
\label{fig:Fig10nog} 
\end{figure}

\begin{figure}[htbp!]
\includegraphics[width=\columnwidth]{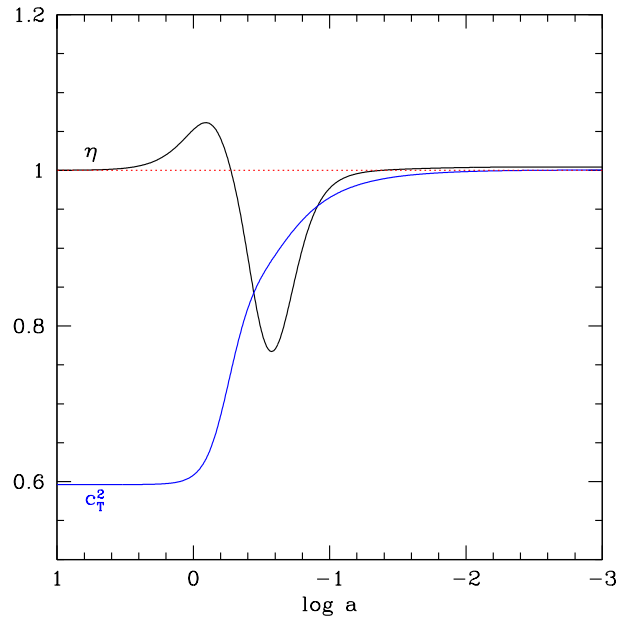} 
\caption{As Fig.~\ref{fig:Fig10nog} but for a derivatively coupled 
Galileon gravity model. Again, at late times the slip 
goes to unity, the general relativity value, but $c_T^2$ does not. 
} 
\label{fig:Fig10g} 
\end{figure}

We solve the coupled equations of motion 
$x'(H,x)$, $H'(H,x)$ (e.g.\ see Eqs.~8, 9 of \cite{appgal}) from standard 
early matter dominated initial conditions 
to the future de Sitter attractor, and use the results in 
Eqs.~(\ref{eq:eta}) and (\ref{eq:ct}). In Fig.~\ref{fig:Fig10nog} we show the 
evolution of the slip and gravitational wave sound speed for the uncoupled 
Galileon gravity case of Fig.~10 from \cite{appgal}. Here $c_{2\dots 5}$ 
are nonzero and the present dark energy density is $\Omega_{\pi,0}=0.72$. 
We clearly see that $\eta=1$ does not imply $c_T^2=1$ during the late time 
de Sitter state, or in the recent past. Similarly this verifies that 
$c_T^2=1$ does not imply $\eta=1$. Thus the lack of scalar deviations 
at some moment does not guarantee the lack of tensor deviations, or the converse. 
The derivative coupled 
Galileon case, also as in Fig.~10 of \cite{appgal}, is shown in 
Fig.~\ref{fig:Fig10g} and again demonstrates that no firm relation exists 
between slip and gravitational wave speed. These cases are free of ghosts 
or instabilities.

Note that while $\eta_{\rm dS}=1$, this does not mean $\geff=G_N$. Indeed 
from Fig.~10 of \cite{appgal} we see that $\geffds$ can be 
substantially greater than unity: $\geff/G_N\approx33$ for the uncoupled case 
and 6.5 for the coupled case. This highlights that $\eta$ is not 
the only ingredient in the modified Poisson equations; observationally we 
also care about the absolute gravitational strength, e.g.\ $\geffs$ entering 
growth of structure or $(1+\eta)\,\geffs$ entering gravitational lensing. 
Both growth and lensing must be measured to extract $\eta$. 
(See \cite{12100439} for details on how other cosmological and astrophysical 
quantities enter the observations, and which combinations can be determined.) 

\begin{figure}[htbp!]
\includegraphics[width=\columnwidth]{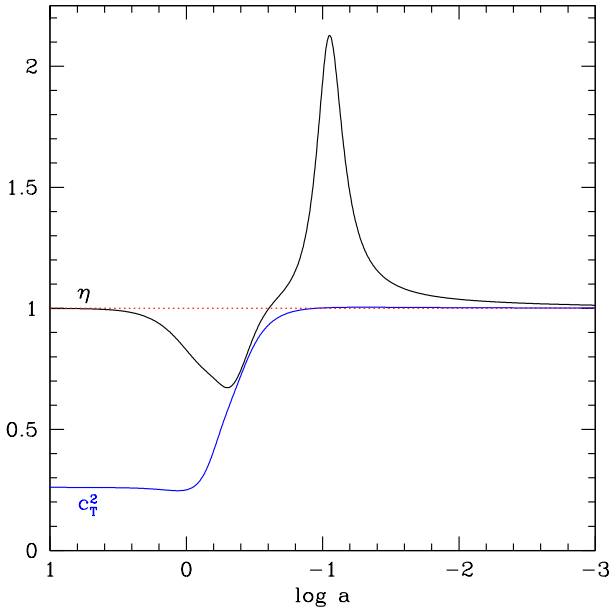} 
\caption{As Fig.~\ref{fig:Fig10nog} but for an uncoupled 
Galileon gravity model where the gravitational strength $\geff$ returns 
to general relativity at late times. Although at late times the slip 
also goes to the general relativity value of unity, $c_T^2$ does not. 
} 
\label{fig:Fig6geff1} 
\end{figure}

To demonstrate that $\geff\ne G_N$ is not responsible for $c_T^2\ne1$, 
despite $\eta=1$, we plot in Fig.~\ref{fig:Fig6geff1} an uncoupled case 
from Fig.~6 of \cite{appgal} where $\geffds=G_N$. Although 
$\eta_{\rm dS}=1$ and $\geffds/G_N=1$, still $c_T^2\ne1$; while the scalar sector 
looks like general relativity with a cosmological constant, the tensor 
sector has a clear deviation.

That conclusion holds for an instantaneous measurement. Let us now 
consider the global case, where $\eta=1$ for all time, and its 
implications for the tensor sector. (Note from Eq.~\ref{eq:ct21} that 
the only way $c_T^2=1$ for all time is for $c_4=c_5=c_G=0$, i.e.\ 
general relativity must hold.) Suppose $\eta=1$ for all time. (We leave 
aside the early universe where $\eta-1\sim\Omega_\pi\ll1$.) Then from 
Eq.~(\ref{eq:eta1}) we see that its right hand side is a polynomial of 
$H$ and $x$. For this to vanish at all times, the coefficients of the 
polynomial (hence the $c_i$) must be zero, so we have general relativity. 

Thus global nondeviation in slip leads to global nondeviation in $c_T^2$ 
(and all other gravitational wave propagation characteristics) for our 
Galileon case. However, if $\eta\ne1$ for some finite period then $c_T^2$ 
will deviate during some (possibly different) period, and if $c_T^2\ne1$ 
for some finite period then $\eta$ will deviate during some (possibly 
different) period.

\section{Conclusions} \label{sec:concl} 

Modified gravity is an active area of exploration on both the theoretical 
and observational fronts. It has the potential to give rise to cosmic 
acceleration, a disconnect between the expansion history and structure 
growth history, deviations from Newton's constant in the gravitational 
strengths entering the modified Poisson equations (changing cosmic growth 
and gravitational lensing), and deviations from general relativity in 
gravitational wave propagation. Detecting these deviations would be 
revolutionary, and if there were a connection between the deviations then 
this might lead to insights about the modified gravity theory. 

We explored whether such a connection necessarily exists between the 
gravitational slip of the scalar metric potentials and the tensor modes, 
as recently suggested by \cite{14067139}. The answer is no and yes. 
Working 
within covariant Galileon gravity, and its extension to derivative coupling 
to the Einstein tensor (as appears in other modified gravity theories), 
we show both analytically and numerically that one can have zero 
deviation in slip at some moment and 
still have a gravitational wave speed different from the speed of light, 
hence a separate signature of breaking general relativity. In particular, 
during the de Sitter attractor of cosmic acceleration or even the incompletely 
matter dominated era, this holds. Similarly, one can have a gravitational 
wave speed equaling the speed of light at some moment, but a deviation 
in the gravitational slip. Only when the effective dark energy density 
makes a negligible contribution, $\Omega_\pi\ll1$, does lack of a deviation 
in one sector necessitate simultaneous general relativistic behavior in the 
other. Globally in time, however, if one deviation ever occurs, the other will 
deviate at some time; and if one deviation never occurs, the other will 
never deviate. 

Taken together, these conditions imply that observations of cosmic large scale 
structure, its growth, evolution, and gravitational lensing effects, are 
complementary to future measurements of gravitational waves. Gravitational 
wave observations thus have a definite role to play in understanding the 
nature of cosmic acceleration and gravity.

\acknowledgments 

I thank Stephen Appleby, Antonio De Felice, Iggy Sawicki, and Shinji 
Tsujikawa for helpful remarks, 
Doyeon Kim for helpful crosschecks, and KASI, IBS, and Kavli IPMU for 
hospitality during part of this work. 
This work has been supported by DOE grant DE-SC-0007867 and the Director, 
Office of Science, Office of High Energy Physics, of the U.S.\ Department 
of Energy under Contract No.\ DE-AC02-05CH11231.

\appendix 

\section{Relation to Gravitational Property Functions} \label{sec:apxalf} 

In \cite{14043713} they describe the gravitational sector in terms of 
four ``property functions'' $\alpha_i$: $\alpha_K$ describing the kinetic 
properties, $\alpha_B$ the braiding of the kinetic and metric terms, 
$\alpha_M$ the running of the Planck mass, and $\alpha_T$ the tensor 
gravitational wave speed. In Galileon gravity, the gravitational wave 
propagation equation is affected only by their speed and the Planck mass 
running. The latter vanishes in the de Sitter limit, however, so we have 
focused here on the gravitational wave speed; our conclusions do not alter 
when considering the other terms. 

In terms of our notation, we can write 
\beqa 
& & \alpha_T=c_T^2-1=\frac{2\kappa_3}{\kappa_4}-1\\ 
& & \alpha_B=\frac{2\kappa_5 x}{\kappa_4}\\ 
& & \alpha_K=\frac{4\kappa_2 x^2}{\kappa_4}\\ 
& & M_*^2=\frac{-\kappa_4}{2}\\ 
& & \alpha_M=\frac{d\ln M_*^2}{d\ln a}=\frac{\kappa'_4}{\kappa_4} \ . 
\eeqa 

The gravitational slip involves two terms contributing to the anisotropic 
stress, from the scalar sector and from the gravitational waves. This 
leaves open the possibility that the terms can cancel under special 
circumstances, and this allows $\eta=1$ in the de Sitter limit despite 
$c_T^2$ deviating from 1 (and hence $\alpha_T\ne0$). Conversely, when 
$c_T^2=1$ at some moment (say in the recent past), the slip can still deviate 
from general relativity because of a contribution proportional to $\alpha_M$. 

Finally, note that \cite{11084242} and \cite{14043713} caution that the 
quasistatic approximation involves not only the Hubble scale but the 
braiding scale. Basically, if the coefficients of the spatial derivative terms 
(Laplacians) in the equations for the metric potentials become small, then 
we can no longer neglect the time derivative terms. The condition for 
continued validity of the quasistatic approximation is 
\beq 
\alpha_B\,\frac{k}{aH}\gg1 \ . 
\eeq 
We have verified numerically that for the cases used in our Figures, 
today $\alpha_B\sim{\cal O}(1)$ and $\alpha_B>0.01$ for $a>10^{-3}$, 
and so the quasistatic approximation holds on sufficiently 
subhorizon observational 
scales.

\section{Galileon Functions} \label{sec:apxkap} 

In Eq.~(\ref{eq:eta}) we use quantities $\kappa_i$ for notational 
simplicity. Here we exhibit them for covariant Galileon gravity allowing 
derivative coupling, following \cite{appgal}. 

\begin{widetext} 
\begin{eqnarray} 
 & & \kappa_{1} = -6 c_{4} {H}^{3}x^{2} \left( {H}' x + {H}
 x' + {{H} x \over 3} \right) + 2c_{\rm G} \left( {H}{H}' x + {H
}^{2} x' + {H}^{2}x \right) +c_{5} {H}^{5} x^{3} \left( 12 
{H} x'  + 15 {H}' x + 3{H} x \right)   \\ 
 & & \kappa_{2} = -{c
_{2} \over 2} + 6c_{3}{H}^{2} x + 3 c_{\rm G} {H}^{2} - 27c_{4} {H}^
{4}x^{2} + 30 c_{5} {H}^{6} x^{3}  \\ 
 & & \kappa_{3} = -1 - {c_{4} \over 2} {H}^{4} x^{4} + c_{\rm G}{H}^{2}x^{2} - 3c_{5} {H}^{5}x^{4}\left({H}x' + {H}'x \right)  \\ 
 & & \kappa_{4} = - 2 + 3c_{4}{H}^{4} x^{4}  -2c_{\rm G}{H}^{2}x^{2} -6c_{5} {H}^{6}x^{5} \\ 
 & & \kappa_{5} = 2c_{3} {H}^{2} x^{2} - 12c_{4} {H}^{4} x^{3}  + 4c_{\rm G}{H}^{2} x + 15c_{5}{H}^{6}x^{4} \\ 
 & & \kappa_{6} =  {c_{2} \over 2} - 2c_{3}  \left( {H}^{2} x' + {H}{H}' x + 2{H}^{2}x \right) + c_{4}  \left( 12{H}^{4}x x' + 18{H}^{3}x^{2}{H}'  + 13 {H}^{4}x^{2} \right)  \\ 
\nonumber & &\hspace{10mm} - c_{\rm G} \left( 2{H}{H}' + 3{H}^{2} \right) - c_{5}  \left( 18{H}^{6}x^{2}x' + 30{H}^{5}x^{3}{H}' + 12{H}^{6}x^{3} \right)  \ . 
\end{eqnarray} 
\end{widetext}


\end{document}